\documentclass[a4paper]{jpconf}
\usepackage{graphicx}
\begin{document}
\title{Fiber laser development for LISA}

\author{Kenji Numata$^{1,2,*}$, Jeffrey R. Chen$^{3}$, Jordan Camp$^{2}$}

\address{$^1$Department of Astronomy, University of Maryland, College Park, Maryland, 20742, USA}
\address{$^2$NASA Goddard Space Flight Center, Gravitational astrophysics branch, Code 663, Greenbelt, Maryland, 20771, USA}
\address{$^3$NASA Goddard Space Flight Center, Laser and electro-optics branch, Code 554, Greenbelt, Maryland, 20771, USA}
\ead{$^*$kenji.numata@nasa.gov}

\begin{abstract}
We have developed a linearly-polarized Ytterbium-doped fiber ring laser with single longitudinal-mode output at 1064nm for LISA and other space applications.
Single longitudinal-mode selection was achieved by using a fiber Bragg grating (FBG) and a fiber Fabry-Perot (FFP).
The FFP also serves as a frequency-reference within our ring laser. Our laser exhibits comparable low frequency and intensity noise to Non-Planar Ring Oscillator (NPRO).
By using a fiber-coupled phase modulator as a frequency actuator, the laser frequency can be electro-optically tuned at a rate of 100kHz.
It appears that our fiber ring laser is promising for space applications where robustness of fiber optics is desirable. 
\end{abstract}

\section{Introduction}

Single frequency fiber laser technology has made great advances over the last ten years and is overcoming limitations of traditional bulk-optics based lasers such as the Non-Planar Ring Oscillator (NPRO).
The NPRO exhibits low frequency fluctuations due to small deformations of bulk crystal that forms the laser cavity \cite{Kane1985}, and has been widely used in low-noise, single-frequency applications.

On the other hand, there is a great interest to develop single frequency fiber lasers.
Compared to a NPRO, a fiber laser offers significant advantages:
1) A fiber laser is virtually alignment free due to the wave-guided laser cavity and pump laser path, and thus more robust against mechanical disturbances;
2) The fiber waveguide maintains single mode and linear polarized laser beam that can be readily coupled into fiber amplifiers;
3) A strong magnet is not needed;
4) A fiber laser is also contamination free due to the closed cavity;
5) It is easier to implement component redundancy in a fiber laser.

The high robustness and efficiency of fiber lasers are particularly attractive for space applications.
It has been proposed for LISA mission \cite{LISA} that a fiber-coupled waveguide phase modulator and a Ytterbium(Yb)-doped fiber amplifier be incorporated in the laser transmitter in order to modulate and amplify the laser.
This makes a fiber oscillator more attractive for its inherent fiber coupled output.

We have developed a Yb-doped fiber ring laser that emits linearly-polarized, single longitudinal-mode, and continuous-wave light at 1064nm for space applications such as LISA and GRACE follow-on \cite{Stephens2006}.
This laser was built solely with commercially available components.
Single longitudinal-mode was selected by two filters in series, a fiber Bragg grating (FBG) and a fiber Fabry-Perot (FFP).
We achieved mode-hop free operation and low frequency and intensity noise performances comparable to commercial NPRO.
The optical frequency of the laser can be varied through slow and fast actuators to facilitate frequency stabilization using external references.
Coarse but slow frequency tuning was achieved by changing FFP spacing (and FBG temperature for large change), and fast tuning was enabled through an intra-cavity phase modulator.
It appears that our fiber ring laser is promising for space applications.
The details of this laser are described in the following sections.

\section{Experimental setup}

\begin{figure}
\begin{minipage}{22pc}
\includegraphics[width=22pc]{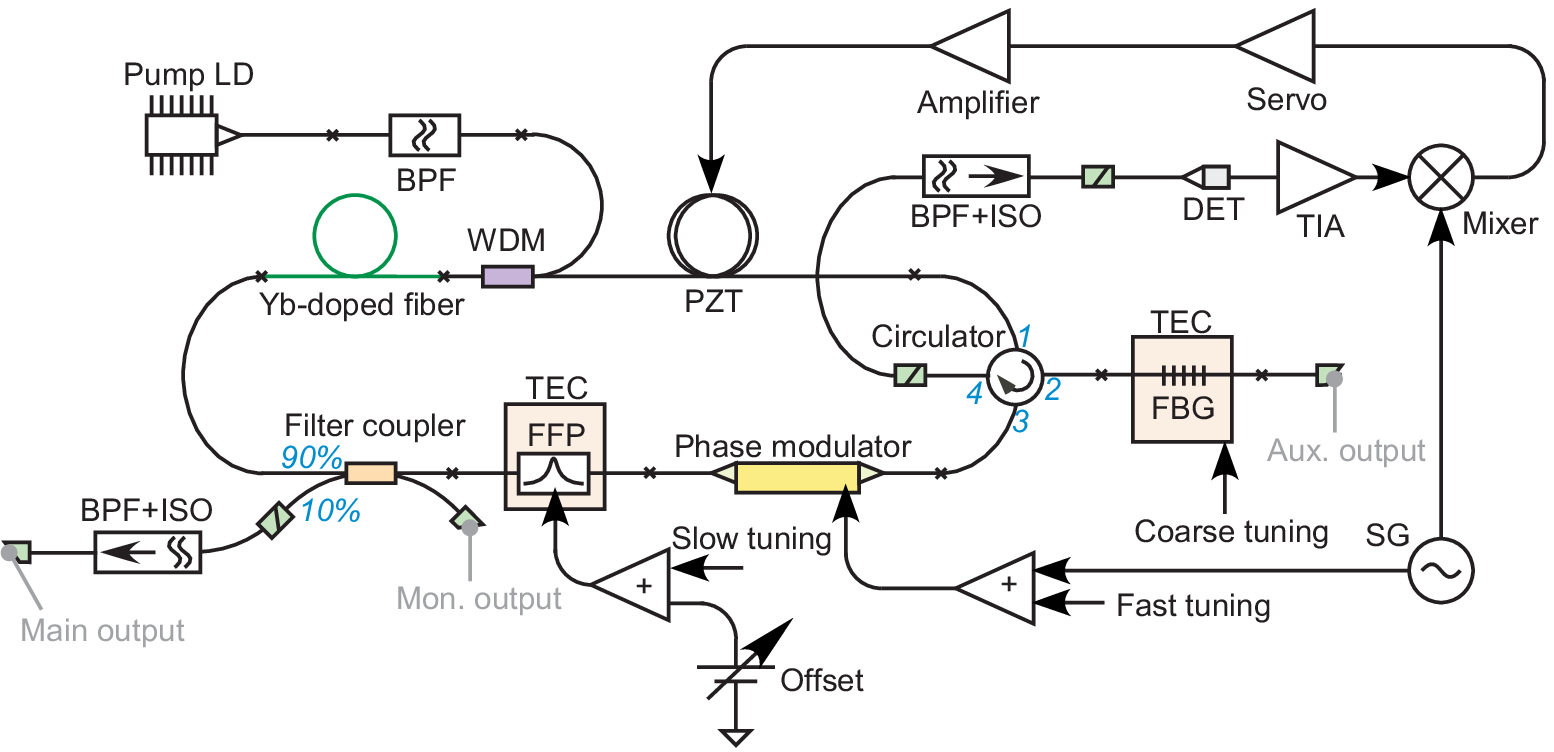}
\caption{\label{RingLaserConfiguration}Ring laser configuration. WDM: wavelength division multiplexing coupler, FBG: fiber Bragg grating, FFP: fiber Fabry-Perot, BPF: band-pass filter, ISO: isolator, DET: detector, TIA: transimpedance amplifier, TEC: thermo electric cooler, SG: signal generator.}
\end{minipage}\hspace{1pc}%
\begin{minipage}{14pc}
\includegraphics[width=14pc]{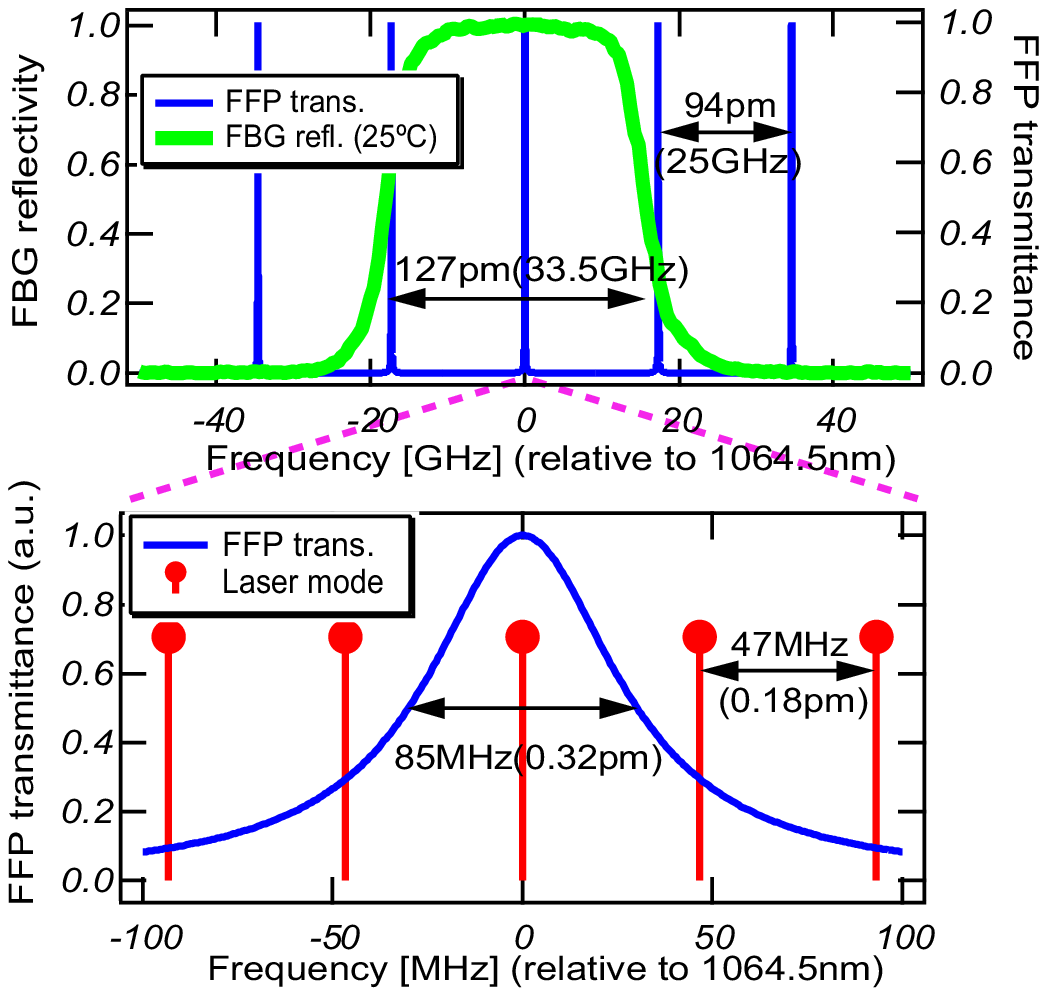}
\caption{\label{ModeSelection}Mode selection. Bottom figure magnifies central region of top figure.}
\end{minipage} 
\end{figure}

Figure \ref{RingLaserConfiguration} shows our ring laser configuration.
The Yb doped gain fiber in the ring cavity was core pumped by a laser diode (LD) through a wavelength division multiplexing (WDM) coupler.
Single longitudinal mode selection was achieved by cascading a FBG and a FFP.
This configuration is similar to earlier work done at 1.5${\rm \mu}$m range using Erbium-doped, non-polarization maintaining (PM) fibers \cite{Inaba2002, Cheng2008}.
By using PM Yb fiber and PM components, our ring laser produces stable and linearly-polarized output at 1064nm.
We used different control schemes and introduced fast frequency tuning to minimize noise.

\subsection{Filters and mode selection}

Figure \ref{ModeSelection} illustrates how single longitudinal mode was selected.
We used a FBG as a coarse filter to select lasing wavelength out of Yb's wide gain bandwidth that spans over 100nm around 1030nm.
The FBG was written on the slow axis of a PM980 fiber and its peak reflectivity was over 99\%.
The center wavelength of its reflection was 1064.5nm at room temperature and the reflection bandwidth was 0.127nm (33.5GHz).
The FBG was inserted into a temperature-controlled copper block for coarse wavelength tuning.
The FBG was spliced to port 2 of a 4-port circulator, so that the FBG was used in its reflection mode.
The light going into port 1, 2, and 3 comes out from port 2, 3, and 4, respectively.
The unidirectional operation of this fiber ring laser was achieved by this 4-port circulator.

The FFP was used as the second filter to select one of the longitudinal modes within the FBG bandwidth.
In FFP, a Fabry-Perot (FP) cavity is formed between two PM980 fiber ends.
Free spectral range (FSR) and finesse of the FP cavity was 25GHz and 290, respectively, and corresponding bandwidth was $\sim$85MHz.
The FFP's 25-GHz FSR restricted the lasing to the center of the FBG bandwidth.
The 85-MHz FFP's BW then selected one longitudinal mode of the $\sim$4.4-m laser cavity, whose FSR was about 47MHz.

\subsection{Control systems}

In order to keep the single-mode oscillation, the lasing longitudinal mode and the FFP resonance must be aligned.
Pound-Drever-Hall (PDH) technique \cite{Hall1981} was used to lock the cavity to the FFP.
The reflected light from FFP is passed though a intra-cavity lithium-niobate phase modulator that phase modulates the light at 80MHz, and is directed through port 4 of the circulator to a fiber-coupled detector.
The detected signal is demodulated using a mixer.
An isolator with integral bandpass filter is placed in front of the detector to remove amplified spontaneous emission (ASE) from the laser. 

Once lasing is achieved, the demodulated signal at the mixer represents the difference between the laser frequency and the FFP resonance.
The signal was filtered and fed back to a piezo actuator (PZT) around which a section of the ring cavity fiber was coiled, forcing laser frequency to follow the FFP resonance.
Thus, the FFP serves as a frequency reference in this control scheme.
Control bandwidth of this loop was about 1kHz.
Temperature of the FFP was actively stabilized by a thermo electric cooler.
The phase modulator was used also for tuning the laser frequency by changing its optical length with an applied voltage. 
The output intensity of the laser was actively stabilized by monitoring the main output and by controlling the pump current.

\subsection{Pump source and gain media}

The pump LD was single mode, PM, fiber-coupled, and single-longitudinal-mode laser.
The output wavelength was internally stabilized to 976nm, where our gain fiber had maximum absorption.
The pump light was coupled into the cavity through a PM WDM coupler after passing through a narrow-band filter at 976nm.
The filter prevented the ASE and 1064nm laser from reaching the pump LD.

The gain fiber was a double-cladding, single mode, PM, Yb-doped fiber.
We used it as a single-clad fiber, pumping its 6-${\rm \mu}$m core.
The small signal absorption of the core was 1200dB/m at 976nm and the length of the gain fiber was about 40cm.

\begin{figure}
\begin{minipage}{18pc}
\includegraphics[width=18pc]{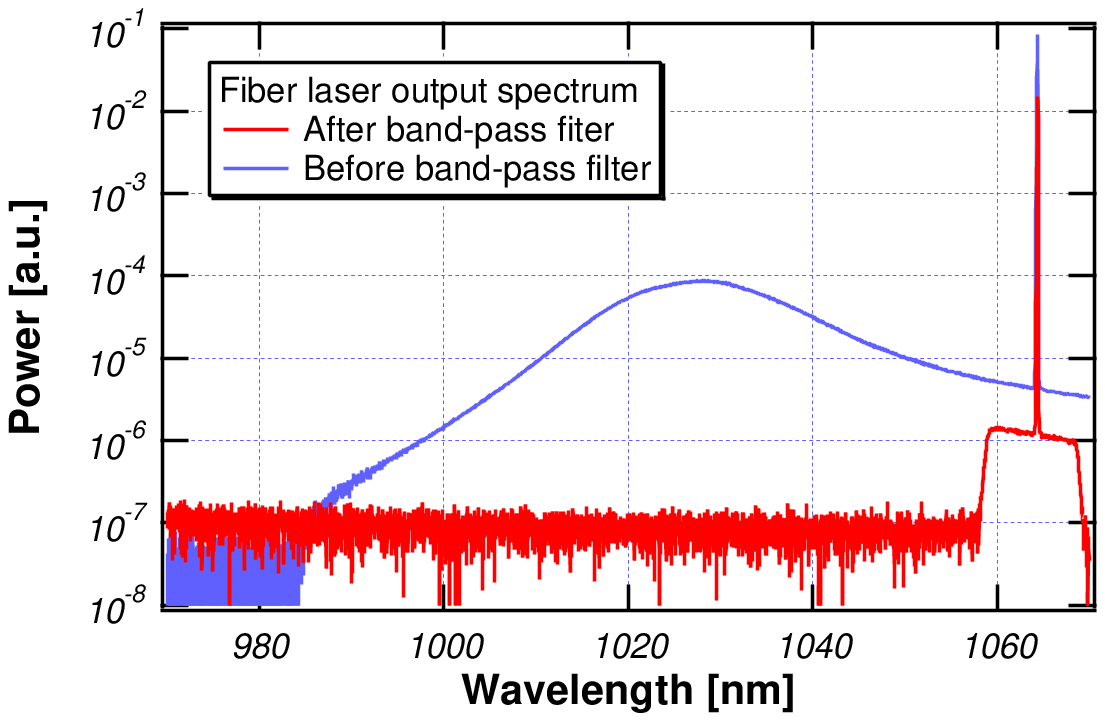}
\caption{\label{OpticalSpectrum}Output optical spectra. ASE is filtered out by a band-pass filter.}
\end{minipage}\hspace{2pc}%
\begin{minipage}{18pc}
\includegraphics[width=18pc]{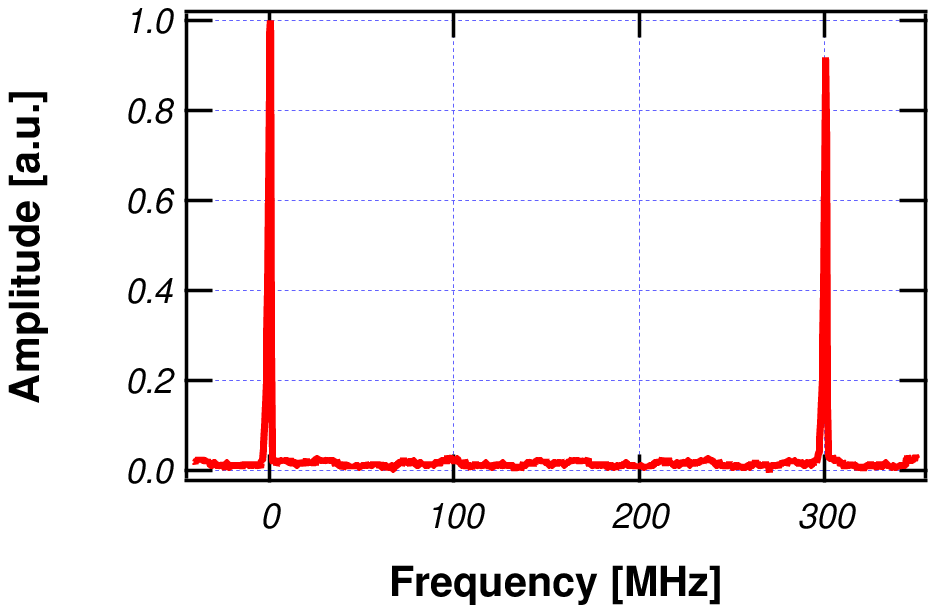}
\caption{\label{LaserMode}Single longitudinal laser mode measured by a scanning Fabry-Perot. Identical mode showed up twice as two peaks due to 300-MHz FSR of the scanning FP.}
\end{minipage} 
\end{figure}

\subsection{Other laser components}

A PM filter coupler was used as an output coupler of the laser.
10\% of light was extracted from the laser cavity, and then was filtered by an isolator with integrated ASE filter.
The location of this output coupler and the coupling ratio were not optimized.

Polarization parallel to fast-axis was blocked in the circulator and isolators.
This prevented lasing along fast axis and improved polarization extinction ratio.

\section{Experimental results}

\subsection{Output optical power and spectrum}

Limited by commercially available components, our experiment was intended to prove the design concept and was by no means optimized.
The excessive insertion loss in the FFP (4.3dB), the phase modulator (2.6dB), and the circulator (4.9dB) resulted in high pump threshold of $\sim$400mW and an output power of $\sim$0.2mW under maximum available pump power of $\sim$600mW.
Placing the output coupler after lossy components and the low coupling ratio (10\%) also contributed to the low efficiency.
The output polarization extinction ratio was better than 20dB.

Figure \ref{OpticalSpectrum} shows output optical spectrum.
The ASE component centered around 1030nm was filtered out by the ASE filter integrated with the output isolator. 

Figure \ref{LaserMode} shows detailed optical spectrum measured by a scanning Fabry-Perot cavity with 300-MHz FSR.
It can be seen that the fiber laser oscillates in single longitudinal mode and the spectrum linewidth was below $\sim$1MHz (resolution limited).
The control system prevented mode-hopping. 

\subsection{Frequency tuning}

\begin{figure}
\begin{minipage}{18pc}
\includegraphics[width=18pc]{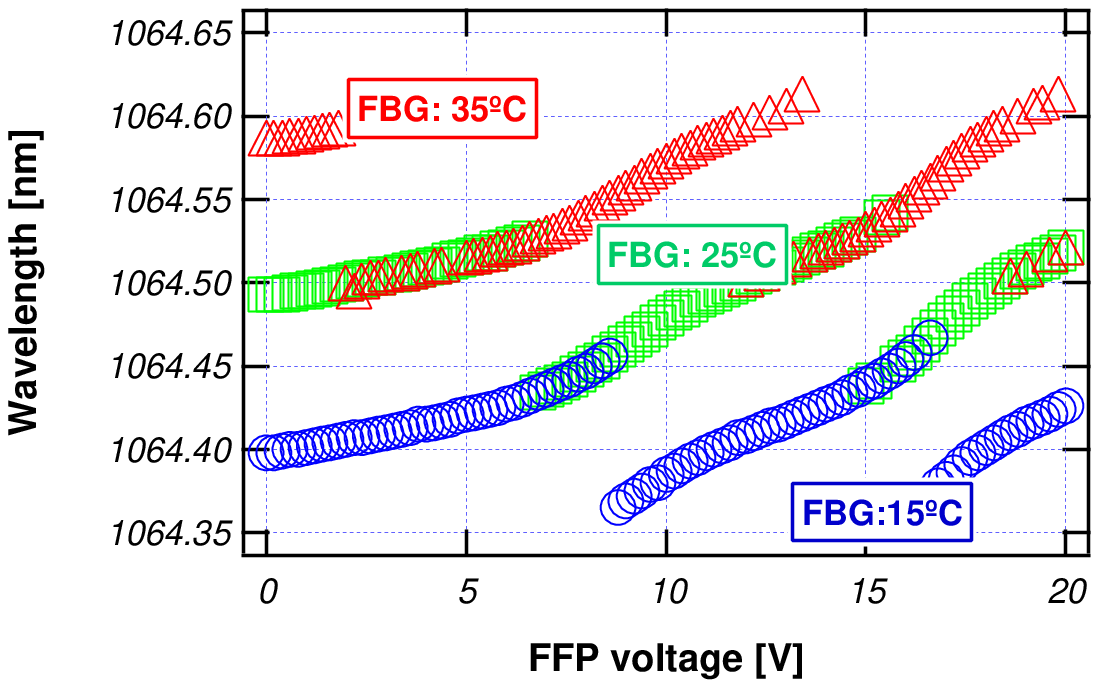}
\caption{\label{TemperatureTuning}Wavelength tuning by FBG temperature and FFP spacing. Results with three different FBG temperatures are shown.}
\end{minipage}\hspace{2pc}%
\begin{minipage}{18pc}
\includegraphics[width=18pc]{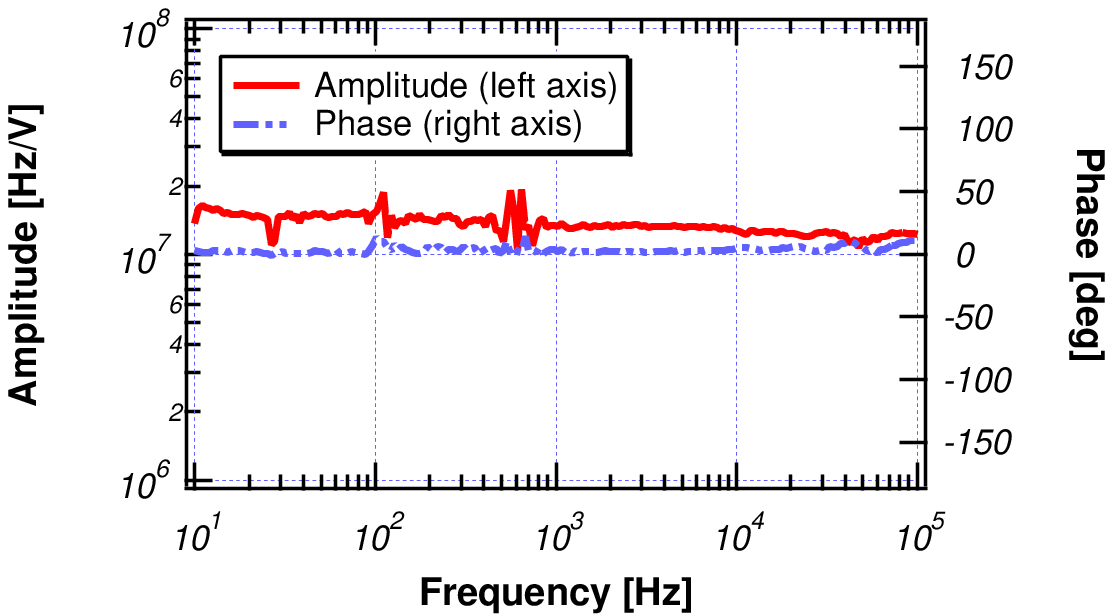}
\caption{\label{PMtuning}Frequency tuning transfer function of the phase modulator used in our ring laser.}
\end{minipage} 
\end{figure}

Coarse wavelength (frequency) tuning was achieved by changing FFP spacing, and FBG temperature for long range tuning.
Figure \ref{TemperatureTuning} shows result of such coarse tuning.
The temperature was tuned between 15$^{\rm \circ}$C and 35$^{\rm \circ}$C, resulting in a center wavelength shifting of 0.25nm (66GHz).
The FFP spacing was changed by varying the voltage applied to the PZT stage of the FFP.
The wavelength can be fast tuned through the FFP PZT in a 1kHz bandwidth. 

Optical length of the cavity can also be fast tuned by varying the voltage applied to the phase modulator.
We merged the modulation signal and the fast tuning signal with a wide-bandwidth operational amplifier.
As shown in Fig.\ref{PMtuning}, the transfer function of such frequency tuning remains flat within the 100kHz measurement range.
The phase modulator enables tuning much faster than commercial NPROs and fiber lasers, in which mechanical deformation is used as a method to change cavity lengths \cite{Trobs2009}.

\subsection{Frequency and intensity noise}

\begin{figure}
\begin{minipage}{18pc}
\includegraphics[width=18pc]{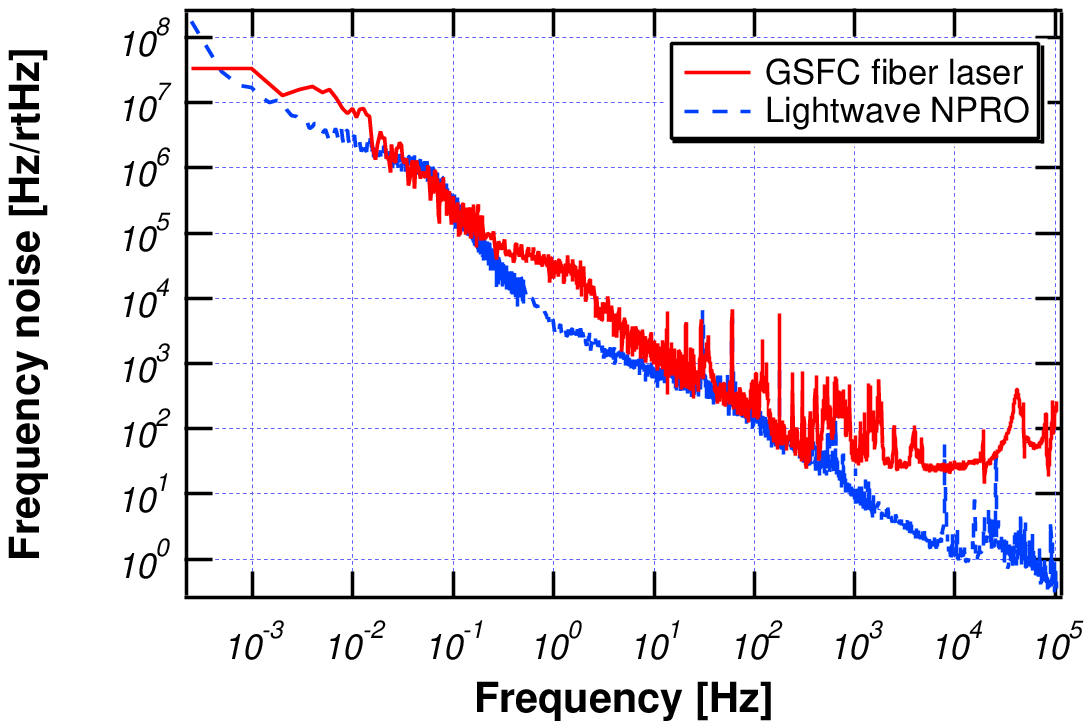}
\caption{\label{FrequencyNoise}Frequency noise of fiber laser and NPRO. Below 10Hz, the measurements were done by taking beatnotes against another NPRO.}
\end{minipage}\hspace{2pc}%
\begin{minipage}{18pc}
\includegraphics[width=18pc]{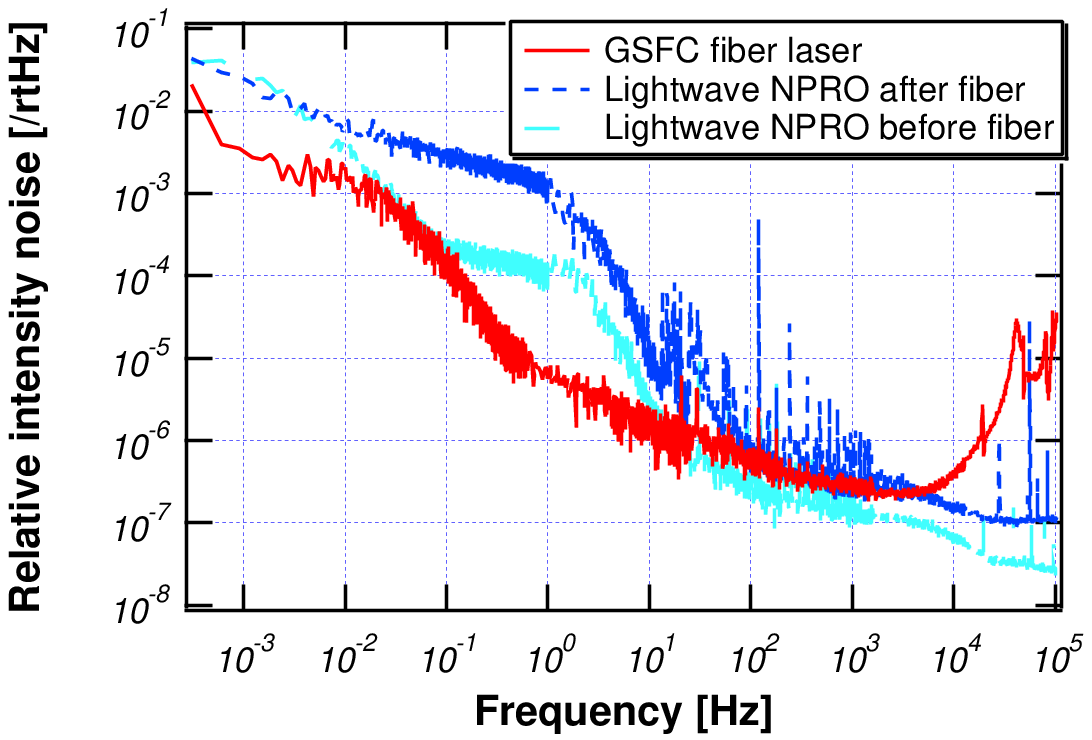}
\caption{\label{IntensityNoise}Relative intensity noise of fiber laser and NPRO. All were measured with internal intensity stabilization turned on.}
\end{minipage} 
\end{figure}

Figure \ref{FrequencyNoise} shows the frequency noise spectrum of our fiber laser in comparison with that of commercial NPRO laser from Lightwave \cite{Lightwave}.
Below 10Hz, the frequency noise were measured by taking beatnotes between a NPRO and the fiber laser and between two NPROs, respectively.
Above 10Hz, the frequency noise were measured with a fiber Mach-Zehnder interferometer with asymmetric arms.

Below 1kHz, our fiber laser had comparable frequency noise to the NPRO.
The frequency noise of our fiber laser was lower around 0.1Hz, and the measured beatnote noise was limited by the NPRO.
Our fiber laser exhibited frequency noise peaks around 1kHz due to acoustic and electronic noise.
Above 1kHz, our fiber laser had larger frequency noise than the NPRO, due to the relaxation oscillation in our fiber laser.

Figure \ref{IntensityNoise} shows relative intensity noise of the fiber laser and the NPRO.
The NPRO had larger intensity noise once its output was fiber-coupled, due to beam pointing fluctuations.
Below 10Hz, our fiber laser had about 10 times lower intensity noise than the fiber-coupled NPRO.
Above 10kHz, our fiber laser had larger intensity noise due to the relaxation oscillation at $\sim$40kHz.
Similar relaxation oscillation frequency has been observed in Yb fiber lasers \cite{Huang2005}.

We also connected output of our fiber laser to a dual stage, core pumping, PM, Yb-doped fiber amplifier and stabilized its output intensity by controlling pump current of the amplifier.
We confirmed that the relative intensity noise can be stabilized down to $\sim 10^{-4}{\rm /\sqrt{Hz}}$ level below 0.1Hz, which satisfies LISA's low-frequency intensity noise requirement.  

\section{Discussion}

In order to improve efficiency of the fiber laser, we are experimenting with simpler optical configurations that minimize optical losses.
One way to reduce loss is to remove the output filter coupler, and use the leakage through the FBG as the laser output.
A low-reflectivity FBG will increase the output coupling ratio.
The optical loss in the FFP was one of the main causes of low efficiency.
In our next steps, the FFP will be replaced by a phase-shifted FBG or fiber-coupled solid etalon, which should have smaller insertion loss and higher dimensional stability.
We also expect to improve noise performance by these modifications.
We can also produce more output power by adding another pump laser diode to boost the pump power.
Frequency stabilization of our fiber ring laser to optical cavity or iodine is also planned. 

For space applications, it is important to have internal redundancy, especially for the pump LDs.
In the case of single-mode, PM, core-pumping LD that we used, additional pump LDs can be easily added by polarization combining and by pumping in both directions without introducing large insertion losses. 

Reliability of our fiber laser optical components are planned to be tested in 2010 in collaboration with Lucent Government Solutions (LGS), including vibration, thermal cycling, and radiation.
At LGS, environmental tests of 2-W Yb fiber amplifier components for use in LISA have started.
We expect to complete these fiber laser and amplifier tests in a year, to help identify the final laser configuration.

\section{Summary}

We developed a fiber ring laser for space applications including LISA.
Our fiber laser offers comparable frequency and intensity noise to an NPRO, but also faster frequency tuning, higher polarization extinction ratio, inherently fiber-coupled output, and open architecture in which all optical components are commercially standard and testable.
Future work will include solving problems associated with high insertion losses by using simpler optical configuration and different narrow-band filters.
Space qualification has been started to verify robustness of fiber optic components.
We believe the evolving technologies of fiber lasers and amplifiers will become the lasers of choice for space interferometer applications within the time frame of LISA, and this work is an important initial step for it.

\end{document}